\documentclass[a4paper, 12pt]{article}

\usepackage[]{graphicx}
\usepackage{caption}
\usepackage{subcaption}

\usepackage{epsf,amsmath,bbold,amsfonts,stmaryrd}
\usepackage{mathrsfs}
\usepackage{tikz}
\usetikzlibrary{arrows,matrix,positioning}
\usepackage{appendix}
\usepackage{amssymb}
\usepackage{float}
\usepackage{color} 
\usepackage{dsfont}
\usepackage{bm}
\usepackage[colorlinks]{hyperref}
\hypersetup{pageanchor=false,citecolor=red,urlcolor=red}
\usepackage{indentfirst}
\usepackage{slashed}
\usepackage[numbers,square,comma,compress,merge]{natbib}

\usepackage[skip=2pt,font=small]{caption}

\def\a{\alpha}

\def\f{\frac}
\def\g{\gamma}

\def\p{\partial}

\def\s{\sigma}

\def\be{\begin{equation}}
\def\ee{\end{equation}}

\def\bea{\begin{eqnarray}}
\def\eea{\end{eqnarray}}

\def\ba{\begin{array}}
\def\ea{\end{array}}

\def\bc{\begin{center}}
\def\ec{\end{center}}

\def\bl{\begin{flushleft}}
\def\el{\end{flushleft}}

\def\br{\begin{flushright}}
\def\er{\end{flushright}}

\def\bi{\begin{itemize}}
\def\ei{\end{itemize}}

\def\bt{\begin{tabular}}
\def\et{\end{tabular}}

\newtheorem{question}{Question}
\def\bq{\begin{question}}
\def\eq{\end{question}}

\newtheorem{definition}{Def}
\def\bd{\begin{definition}}
\def\ed{\end{definition}}

\newtheorem{answer}{Answer}
\def\ban{\begin{answer}}
\def\ean{\end{answer}}

\newtheorem{possibleanswer}{Possible answer}
\def\bpa{\begin{possibleanswer}\normalfont}
\def\epa{\end{possibleanswer}}

\newtheorem{theorem}{Theorem}
\def\bth{\begin{theorem}}
\def\eth{\end{theorem}}

\begin{document}

\begin{titlepage}
\vspace*{-2.5cm}
\br
{\small LMU--ASC~67/20}
\er

\begin{center}
\bf \Large{Islands in Linear Dilaton Black Holes}
\end{center}

\begin{center}
\textsc {Georgios
  K. Karananas,$^{\star}$~Alex Kehagias,$^\dagger$~John Taskas\,$^{\dagger}$}
\end{center}

\begin{center}
\it {$^\star$Arnold Sommerfeld Center\\
Ludwig-Maximilians-Universit\"at M\"unchen\\
Theresienstra{\ss}e 37, 80333 M\"unchen, Germany\\
\vspace{.4cm}
$^\dagger$Physics Division\\
National Technical University of Athens\\
15780 Zografou Campus, Athens, Greece\\
}
\end{center}

\begin{center}
\small
\texttt{
georgios.karananas@physik.uni-muenchen.de}\\
\texttt{
kehagias@central.ntua.gr}\\
\texttt{taskas@mail.ntua.gr}
\end{center}

\vspace{1cm}

\begin{abstract}

We derive a novel four-dimensional black hole with planar horizon that asymptotes to the linear dilaton 
background.  The usual growth of its entanglement entropy before Page's time is established. After that, emergent islands modify to a large extent  the entropy, which becomes  finite and is saturated by its Bekenstein-Hawking value in accordance with  the finiteness of the  von Neumann entropy of eternal black holes. We  demonstrate that viewed from the string frame, our solution is the two-dimensional Witten black hole with two additional free bosons. We generalize our findings by considering a general class of linear dilaton black hole solutions at a generic point along  the $\sigma$-model renormalization group (RG) equations. For those, we observe that the entanglement entropy is ``running'' i.e. it is changing along the RG flow  with respect to the two-dimensional worldsheet length scale. At any fixed moment before Page's time the aforementioned entropy increases towards the infrared (IR) domain, whereas the presence of islands  leads the running entropy to decrease towards the IR at later times. Finally, 
we present a four-dimensional charged black hole that asymptotes to the linear dilaton background as well. We compute the associated entanglement entropy for the extremal case and we find that an island is needed in order for it to follow the Page curve.

\end{abstract}

\end{titlepage}

\section{Introduction and motivation}

 Even though the ``problem'' with reproducing Page's curve has explicitly been addressed and solved in a full microscopic theory, the N-portrait~\cite{Dvali:2011aa,Dvali:2012rt,Dvali:2012wq,Dvali:2015aja}, a self-consistent semiclassical tool capable of yielding the correct entropy of a system coupled to a gravitational background was missing. Hitherto there have been important developments towards this direction; for a nice review see~\cite{Almheiri:2020cfm} and~\cite{Penington:2019npb,Almheiri:2019psf,Almheiri:2019hni,Almheiri:2019yqk,Chen:2019uhq,Almheiri:2019psy,Penington:2019kki,Almheiri:2019qdq,Chen:2019iro,Bhattacharya:2020ymw,Gautason:2020tmk,Anegawa:2020ezn,Hashimoto:2020cas,Hartman:2020swn,Hollowood:2020cou,Krishnan:2020oun,Alishahiha:2020qza,Banks:2020zrt,Geng:2020qvw,Chen:2020uac,Chandrasekaran:2020qtn,Li:2020ceg,Bak:2020enw,Bousso:2020kmy,Hollowood:2020kvk,Krishnan:2020fer,Engelhardt:2020qpv,Karlsson:2020uga,Gomez:2020yef,Chen:2020jvn,Hartman:2020khs,Balasubramanian:2020coy,Balasubramanian:2020xqf,Sybesma:2020fxg,Chen:2020hmv,Ling:2020laa,Bhattacharya:2020uun,Marolf:2020rpm,Hernandez:2020nem,Matsuo:2020ypv,Goto:2020wnk,Akal:2020twv,Basak:2020aaa,Caceres:2020jcn,Raju:2020smc,Manu:2020tty,Deng:2020ent} for a non-exhaustive list of references.

More specifically, it has been realized that the proper way to semiclassically compute the  entropy of a black hole's Hawking radiation is to take into account the effect of the so-called ``island'' configurations; these regions are located near the horizon of the black hole (either inside or outside), their boundaries correspond to~\emph{extremal surfaces},\footnote{In the sense that they extremize the generalized entropy functional, see below.} and are completely disconnected from the reservoir where the Hawking radiation resides. Nevertheless, they contribute non-trivially to the latter's entropy. The radiation, although sufficiently far from the black hole such that gravitational effects can be neglected, it is actually entangled with the fields in the vicinity of the black hole---an aftermath of its gravitational origin. 

Intuitively, the island(s) play the role of an effective clock that breaks the time translation invariance and thus the self-similarity of the system. This is what forces the entropy to not grow with time in perpetuity but rather undergo a phase transition around Page time $t_{\rm Page}$ and follow a Page curve. Note that islands can be thought of as a ``geometric manifestation'' of an inherent quantum clock by way of inner entanglement and depletion of gravitons in the N-portrait language. The breaking of self-similarity, actually a prediction of this theory, is due to these effects and becomes maximal after the quantum break-time, which is equal to half-decay, and thus, to Page's time.\footnote{We thank G.~Dvali for explaining these issues to us.}

Of course, being able to compute the entropy of Hawking radiation  
does not imply that something robust may be said about information and the associated so-called ``paradox.'' Actually, this ``paradox''  is most probably not there to start with, since it is related to the assumption that the evaporation of the black hole is an exactly thermal process~\cite{Dvali:2011aa,Dvali:2012rt,Dvali:2012wq,Dvali:2015aja,Dvali:2012en,Dvali:2013lva,Dvali:2013eja}. As argued in these works, Hawking's analysis is incomplete and should at best be taken with a pinch of salt: any deviations from thermality should not be neglected since these are expected to be power-law suppressed by inverse powers of the entropy. This is a general argument that does not resort to any explicit quantum theory of gravity. Therefore, in order to conclude about information at all, it would require to take into account corrections not visible in any semiclassical approximation. 
However, even being able to semiclassically reproduce the expected from unitarity time evolution of the black hole's entanglement entropy, i.e. its Page curve, is an  important breakthrough. 

Roughly, the method to deduce the Page curve is the following. First, one introduces a generalized entropy functional that contains two pieces:~\emph{i)}~the von Neumann or fine-grained entropy of the radiation plus the island;~\emph{ii)}~the Bekenstein-Hawking entropy of the island, i.e. $\text{Area(Island)}/4G_N$, with $G_N$ the Newton constant.\footnote{We work in units $c=\hbar=1$.} Then, one looks for all possible saddle points of this functional; these correspond to the island configurations mentioned above.  The generalized entropy evaluated on top of the saddle that minimizes its value is identified with the entanglement entropy of the system. 

Let us clarify that although the procedure we just outlined was at first suggested in the context of holography and AdS/CFT by Ryu and Takayanagi~\cite{Ryu:2006bv}, see also~\cite{Hubeny:2007xt,Lewkowycz:2013nqa,Barrella:2013wja,Faulkner:2013ana,Engelhardt:2014gca} and~\cite{Penington:2019npb,Almheiri:2019psf,Almheiri:2019hni} for further developments, it is actually applicable to any system coupled to a gravitational background. This is further substantiated by the nontrivial fact that in all the examples studied in the literature so far, the entanglement entropy turns out to follow the Page curve when the aforementioned prescription is employed. 

Even though  these considerations  initially focused on two-dimensional systems, it is by now clear that the ``island rule'' extends well beyond two-dimensional spacetimes, e.g.~\cite{Almheiri:2019psy,Hashimoto:2020cas,Alishahiha:2020qza,Geng:2020qvw}. It also saves the day for the higher dimensional Schwarzschild black hole solution, by rendering its entanglement entropy unitary; this was studied in~\cite{Hashimoto:2020cas} for pure Einsteinian gravity in $D\ge 4$ dimensions, while~\cite{Alishahiha:2020qza} generalized these findings when higher derivative curvature invariants are included in the gravitational action. 

It is thus timely and desired to understand whether islands are behind the sensible behavior of the entanglement entropy of other black hole geometries. 
One suitable arena to test  the island hypothesis is provided by the four-dimensional generalization of the linear dilaton theory, since as we demonstrate, the model accommodates a black hole geometry. It turns out that the proper behavior
of its entanglement entropy after the Page time is reproduced only when an island is included here as well; not accounting for its explicit contribution in the entropy 
yields a result which grows with time and will eventually be in conflict with the assumption of a unitary evolution for the system.

Interestingly, when written in  string frame, the temporal-radial part of the linear dilaton black hole coincides with the Witten cigar solution~\cite{Witten:1991yr}. We repeat the computation of the entanglement entropy and find that the island prescription reproduces the same Page curve, as expected. 
Since the model under consideration is string-inspired---it corresponds to a fixed point of the two-dimensional sigma model---we construct a general class of black holes that satisfy the RG equations, without vanishing beta functions for the metric \& dilaton. For those, their corresponding horizons ``run,'' in the sense that they depend on the worldsheet length scale $\ell$ and approach zero as we move towards the IR. In turn, the associated entanglement entropy acquires a nontrivial dependence on $\ell$ and thus changes under the RG flow. We find that it is not a monotonic function of $\ell$, but rather exhibits a discontinuous behavior in the following sense: the entropy at any moment of time  before $t_{\rm Page}$ grows towards the IR. Around Page time, when the entropy undergoes a phase transition due to the emergence of an island, it starts decreasing as $\ell$ approaches infinity. 

Finally, we couple the four-dimensional linear dilaton model to the electromagnetic field. In this case the equations of motion admit a charged black hole solution that vaguely speaking corresponds to the analog of a planar Reissner-Nordstr\"om solution for the linear dilaton theory. We find the conditions for extremality and for simplicity we confine ourselves to such configurations. We redo the entropy computation and also find that the island is a necessary ingredient for the ``reasonable'' behavior of the system after Page's time. 

This paper is organized as follows. In Sec.~\ref{sec:linear_dilaton_Einstein}, we
construct in details the black hole solution in the Einstein frame and discuss
its properties. In
Sec.~\ref{sec:entangl_entropy}, we spell out the main steps for
computing the entanglement entropy and apply it
to the linear dilaton black hole solution (in  the Einstein frame), without and with an island contribution. In Sec.~\ref{sec:linear_dilaton_string}, we move to the string frame
and demonstrate that the temporal-radial part of the solution actually
corresponds to the well known Witten black hole. We repeat the computation of the entropy and show that the results are qualitatively the same in the two frames. In Sec.~\ref{sec:running}, we turn our attention to the more general ``running horizon'' black holes and find the dependence on the entropy on the RG scale. In Sec.~\ref{sec:charged}, we present the extremal charged linear dilaton black hole and its associated entropy. We conclude in
Sec.~\ref{sec:conclusion}.

\section{The linear dilaton black hole---Einstein frame}
\label{sec:linear_dilaton_Einstein}

Consider the following Einstein frame action in $D=4$ spacetime dimensions
\be
\label{eq:dil_act_einst}
I=\frac{1}{16\pi G_N}\int d^4x \sqrt{-g}\bigg(R-\frac{1}{2}\big(\partial \sigma\big)^2+4 k^2 e^{\sigma}\bigg) \ ,
\ee
with $G_N$ the Newton constant, $g=\det(g_{\mu\nu})$, $R$ the Ricci scalar, and $k$ a constant with mass-dimension 1. 
The equations of motion for the theory are obtained by varying the action~(\ref{eq:dil_act_einst}) w.r.t. the metric $g_{\mu\nu}$ and scalar field $\sigma$, which turn out to be 
\bea
&&R_{\mu\nu}-\frac{1}{2}g_{\mu\nu}R=\frac{1}{2}\partial_\mu \sigma\partial_\nu \sigma-\frac{1}{2}g_{\mu\nu}\left(\f{1}{2}(\partial \sigma)^2-4 k^2 e^{\sigma}\right) \ ,\label{eq:einst_eqs_cov} \\
&& \Box \sigma=-4 k^2 e^\sigma \ , \label{eq:dil_eom_cov}
\eea
with $\square= g^{\rho\sigma}\nabla_\rho \nabla_\sigma$ the covariant d'Alembertian.  

Let us focus on static, isotropic vacuum solutions of the form
\bea
&&ds^2 = -B(r) dt^2 + A(r) dr^2 +r^2\left(dx^2+dy^2\right) \ ,\\
&&\sigma\equiv \sigma(r)=\alpha \log(k r) \ , 
\eea
with $\alpha\neq 0$ a real constant to be determined.\footnote{In principle, we have the liberty to shift the dilaton by a constant, say $\s_0$. The black hole solution in this case will be the same as the one we will present in the following, up to an overall rescaling by  $e^{\s_0}$.}~By solving the field equations~(\ref{eq:einst_eqs_cov}) and~(\ref{eq:dil_eom_cov}), we find that 
\be 
A(r) = \frac{1}{1-\frac{r_h^2}{r^2}} \ ,~~~B(r) = 
r^2\left(1-\frac{r_h^2}{r^2}\right) \ ,~~~\alpha=-2 \ ,
\ee
meaning that the four-dimensional theory described by~(\ref{eq:dil_act_einst}) admits the following black hole geometry
\bea
&&d s^2=-r^2\left(1-\frac{r_h^2}{r^2}\right)d t^2+\frac{d r^2}{1-\frac{r_h^2}{r^2}}+r^2\left(d x^2+d y^2\right) \ ,\label{eq:ds40}\\
&&\sigma=-2\, \log (k r)\ . \label{eq:s1} 
\eea
Notice that the above metric for $r_h=0$ (or $r\to\infty$) is not flat but rather approaches
\be
\label{eq:linear1} 
ds^2 =  -r^2dt^2 + dr^2 + r^2(dx^2+dy^2) \ , 
\ee
which is just the linear dilaton, or ``continuous clockwork" geometry~\cite{Giudice1,*Giudice2,KR1,*KR2} (see also appendix~\ref{app:background}).  

In analogy with the Schwarzschild black hole, let us work in terms of the following tortoise coordinate 
\be
r^*=r_h\int \frac{dr}{r\left(1-\frac{r_h^2}{r^2}\right)}=\frac{r_h}{2}\log\left(\frac{r^2}{r_h^2}-1\right) \ ,
\ee
such that the line element~(\ref{eq:ds40}) becomes (after rescaling $t$ to $r_h t$)
\be
\label{eq:tor}
d s^2=-e^{\frac{2r^*}{r_h}}\Big(d t^2-d r^{*2}\Big)+r_h^2\left(1+e^{\frac{2r^*}{r_h}}\right)\left(dx^2+d y^2\right) \ . 
\ee

\begin{figure}[!t]
\centering
\includegraphics[scale=.4]{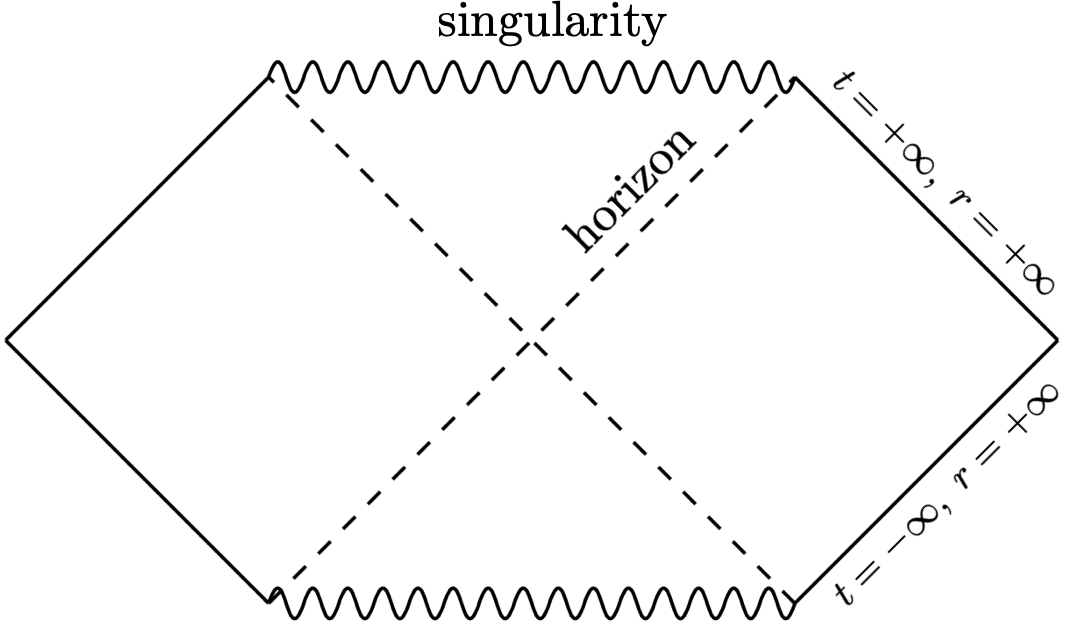}
\caption{Penrose diagram of the black hole. As customary, the curly lines denote the singularity $r=0$, while the dashed ones the horizon located at $r=r_h$. Note that each point of the diagram represents a two-dimensional Euclidean space. Although the conformal factors of the linear dilaton and Schwarzschild black holes are different, their causal structures are the same.}
\label{fig:Penrose_1}
\end{figure}

As usual, $r^*\sim r_h\log (r/r_h)$ for $r\gg r_h$
and $r^*\to -\infty$ for $r=r_h$ where the event horizon lies. To go to the Kruskal coordinates, we define
\be
\label{eq:uvstar}
u^*=t-r^*,~~~v^* =t+r^*,
\ee
so that
\be
r^*=\frac{v^*-u^*}{2}, ~~~t=\frac{v^*+u^*}{2},
\ee
and therefore
\be
r^2-r_h^2=r_h^2 e^{\frac{v^*-u^*}{r_h}}.
\ee
We now introduce 
\be
\label{eq:uv}
u=-r_he^{-\frac{u^*}{r_h}}, ~~~v=r_h e^{\frac{v^*}{r_h}},
\ee
in terms of which the metric is written as
\bea
\label{eq:met_einstein_frame}
&&d s^2=-d u d v+r^2(u,v)\left(d x^2+d y^2\right) \ , \\
&&\sigma=-2\log\big(kr(u,v)\big) \ ,
\eea
where $r(u,v)$ is given by 
\be
\label{eq:rrh}
r(u,v)= \sqrt{r_h^2-u v} \ . 
\ee
Therefore, the singularity is at 
\be
uv=r_h^2 \ ,
\ee
and the event horizon at 
\be
u v=0 \ ,
\ee
i.e., at $r=r_h$ from Eq.~(\ref{eq:rrh}).  
The Penrose diagram of the solution is shown in Fig.~\ref{fig:Penrose_1}. 

\section{Entanglement entropy of the linear dilaton black hole}
\label{sec:entangl_entropy}

In this section we carry out the computation of the entanglement entropy of the linear dilaton black hole's Hawking radiation. In general, computing the entropy for generic quantum field theories is a rather nontrivial task; only in a handful of cases it is possible to actually do so, at least analytically. Fortunately, in what follows the situation simplifies considerably, since we will be effectively working in a two-dimensional setting. This point is pertinent, so let us give some more details.  The dilaton black hole solution we have found---unlike the Schwarzschild metric---admits a two-dimensional planar horizon spanned by the $x$ and $y$ coordinates. This is evident from the explicit form of the metric, cf.~(\ref{eq:ds40}). Therefore, in order to extract finite, meaningful, results, we are in a sense obliged to work in terms of  quantities  defined per unit area if $(x,y)$ parametrize $\mathbb{R}^2$. Alternatively, we may take $(x,y)$ to parametrize a two-torus $T^2$. Consequently,  well known results concerning the entropy of two-dimensional field theories are applicable. 

\subsection{Outline of the procedure}

Let us give an outline of the prescription one has to follow. More details can be found for instance in~\cite{Almheiri:2019qdq,Almheiri:2020cfm} and references therein. 

The starting point is to identify the Hawking radiation of the black hole with an appropriate matter sector coupled to the gravitational theory~(\ref{eq:dil_act_einst}). In two-dimensional considerations this usually corresponds to a free CFT that comprises $N\gg 1$ minimally-coupled massless scalar or fermionic fields with central charge $c\sim N$. We also have to require that $c\ll \frac{r_h^2}{G_N}$. 

\begin{figure}[!h]
\begin{subfigure}{.5\textwidth}
\centering
\includegraphics[scale=.4]{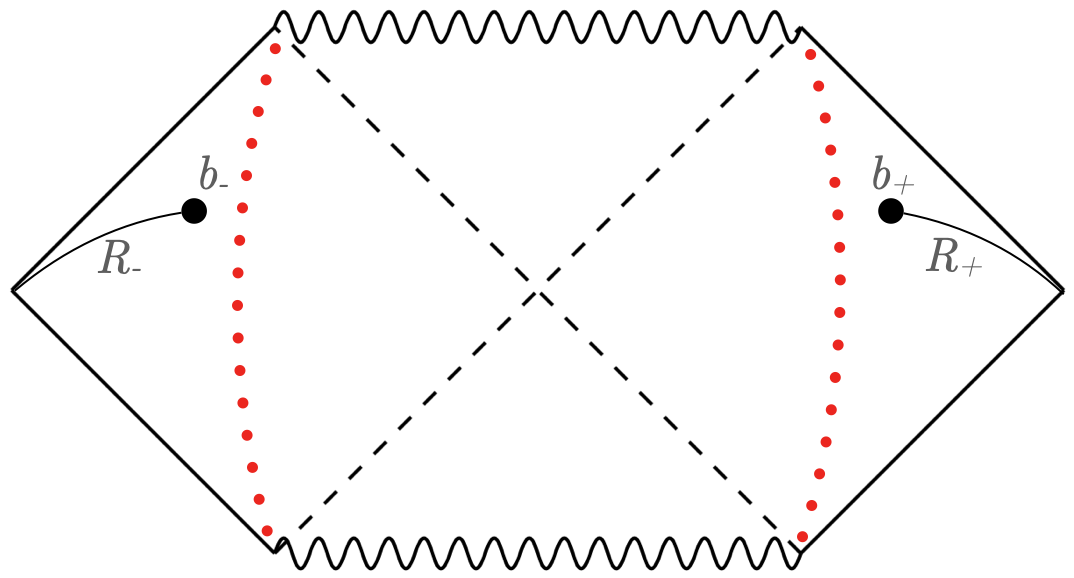}
\caption{}
\label{fig:no_island}
\end{subfigure}
\begin{subfigure}{.5\textwidth}
\centering
\includegraphics[scale=.4]{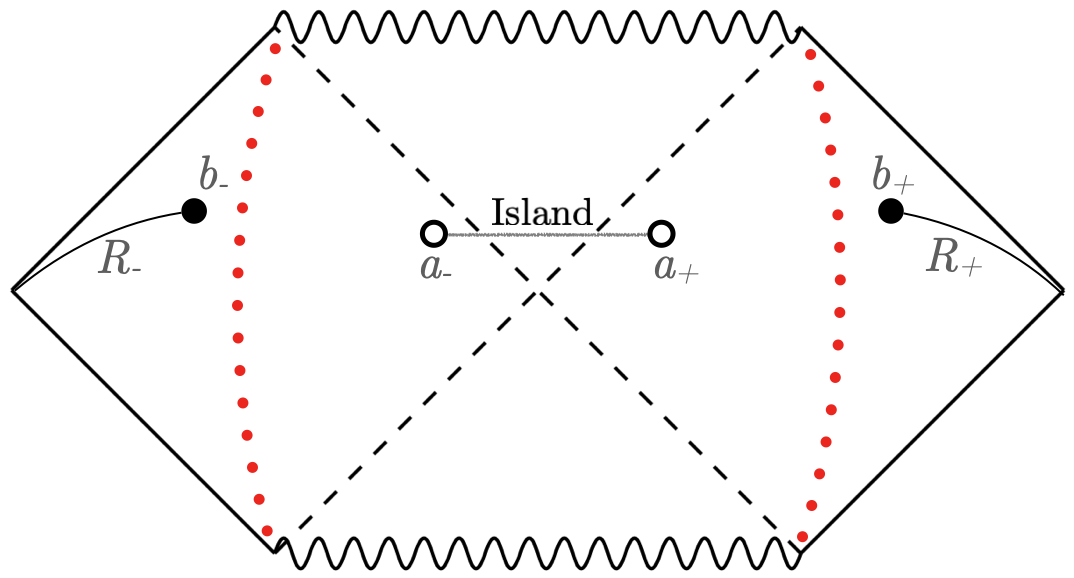}
\caption{}
\label{fig:island}
\end{subfigure}
\caption{(a)~Penrose diagram of the eternal linear dilaton black hole plus the matter fields in the absence of islands. The dotted red lines correspond to the fictitious boundaries of the regions outside of which the effect of the matter sector on the geometry---and vice versa---can be safely neglected.   The black lines are the entanglement regions $R_\mp$ where the radiation lives and $b_\mp$ denoted by blobs their respective  boundaries. Their ranges are $(-\infty,b_-]$ and $[b_+,+\infty)$, respectively; the radiation region is therefore $R=R_-\cup R_+=(-\infty,b_-]\cup [b_+,+\infty)$. ~(b)~The Penrose diagram of the same  system in the presence of an island with boundaries located at $a_\mp$. Note that for the eternal black hole, the island's boundaries lie outside the horizon~\cite{Almheiri:2019yqk,Gautason:2020tmk,Anegawa:2020ezn}.}
\label{fig:Penrose_2}
\end{figure}

It should be very clearly stated that some extra hypotheses are needed in order for the approach to be self-consistent and the results reasonable. 
First, for the matter sector to have negligible backreaction on the background geometry, we have to require that the radiation reservoir is in a region $R$  sufficiently far from the black hole, at least a few Schwarzschild radii away; the (imaginary) surface bounding this region has been marked with red lines in Fig.~\ref{fig:Penrose_2}. To put it differently, the equations of motion of the theory, even in the presence of matter fields, should still accommodate the linear dilaton black hole solution discussed in the previous section. The above has to be supplemented by an  additional assumption: it is not enough that the location of the bath simply be where gravity is weak---the matter system should be nothing more than a ``spectator'' in the sense that it must not gravitate at all. Relaxing the latter requirement yields a trivial Page curve~\cite{Laddha:2020kvp}. 

As we already mentioned in the introductory section, the rule of thumb for computing the entanglement entropy is the following~\cite{Almheiri:2019hni,Almheiri:2020cfm}. One considers a modified entropy $S_{\rm gen}$ which receives a contribution from the Bekenstein-Hawking  entropy of the island, and the von Neumann entropy of the matter sector evaluated on the union of the radiation and the island regions
\be
\label{eq:sgen}
S_{\rm gen}=\frac{\text{Area(Island)}}{4G_N} +S_{\rm matter}\left(R\cup\text{Island}\right) \ . 
\ee
The actual entropy $S$ of the system is found by evaluating $S_{\rm gen}$ on top of all the saddle points (extrema), i.e. the locations of the island, and then singling out the one that yields the minimal value, so that
\be
S= \min\text{ext}\left\{S_{\rm gen}\right\} \ .
\ee
It is the nontrivial interplay between the two contributions in~(\ref{eq:sgen}) that results into $S$ undergoing a phase transition around the Page time and behaving in accordance with unitarity. 

As we will see in details shortly, at early times the generalized entropy~(\ref{eq:sgen}) has no real extrema if we include the contribution stemming from the island. This translates into the quantity being saturated by the matter contribution only. 

If no island is included at later times as well, the black hole will continue emitting an ever-increasing amount of Hawking radiation, meaning that the latter's entropy will asymptotically exhibit linear growth with time. Consequently, the entropy will not follow the Page curve. 

If on the other hand we include an island configuration in the generalized entropy, its presence gives rise to a new (real) saddle point that minimizes the generalized entropy after Page time. When evaluated on top of it, the entanglement entropy of the system turns out to coincide with (twice) the Bekenstein-Hawking entropy $S_{\rm BH}$ of the black hole. 

Before moving on, let us reiterate the main points. At each time, one computes the generalized entropy $S_{\rm gen}$ without and with an island. The entanglement entropy is always  identified with the lowest value of this quantity. The configuration without an island is actually a minimum of the generalized entropy only for times smaller than the Page time. After Page time, one observes that the island is an essential ingredient in order for the entropy to be minimized. 

\subsection{Entropy without island}

By construction, the generalized entropy in the absence of islands coincides with the fine-grained matter entropy, cf.~(\ref{eq:sgen}). For a metric of the form
\be
\label{eq:metric_2D}
ds^2 = - e^{2\rho(u,v)} du dv +r^2(u,v)(dx^2+dy^2) \ ,
\ee
the matter entropy per unit area coincides with the well known two-dimensional formula~\cite{Fiola:1994ir} and reads
\be
\label{eq:entr_formula_noisl}
S_{\rm matter} = \frac{c}{12} \log \Big[ (u_{b_-}-u_{b_+})^2(v_{b_-}-v_{b_+})^2 e^{2\rho(b_+)} e^{2\rho(b_-)}\Big] \ ,
\ee
with $b_{\mp}=(t,r)\equiv(\mp t_b,b)$, the boundaries of the entanglement regions in the left and right wedges of the black hole, see Fig.~\ref{fig:no_island}. Using~(\ref{eq:met_einstein_frame}) and the definition of $u$ and $v$ variables in terms of $t$ and $r$,\footnote{One should be careful to flip the signs of $u$ and $v$ in~(\ref{eq:uv}) when considering points in the left wedge of the Penrose diagram in Fig.~\ref{fig:Penrose_2}, since these correspond to $r<0$.}  we obtain the entropy for the matter fields\,\footnote{Note that for the argument of the logarithm to be dimensionless, we have to divide by appropriate powers of $r_h$.}
\be
\label{eq:smatter_ein}
S_{\rm matter} = \frac {c} {3} \log \left[2 \sqrt{\frac{b^2-r_h^2}{r_h^2}}\cosh \frac{t_b}{r_h}\right]\approx \frac {c} {3} \log \left[\frac{2b}{r_h}\cosh \frac{t_b}{r_h}\right] \ ,
\ee
where we used $b\gg r_h$. From the above expression, we immediately find that without an island, the entropy asymptotically behaves as 
\be
\label{eq:entr_no_island}
S=S_{\rm matter} \sim \frac{c}{3} \frac{t_b}{r_h}  \ ,
\ee
i.e. it grows linearly with time.  It is important to note that in the absence of an island, the time evolution of the entanglement entropy is not compatible with the assumption of unitarity of the system. 

Of course, one can be even more thorough and explicitly show that with an island there is no real solution for $a$ that minimizes the entropy at early times. This translates into the island becoming essential only around and after the Page time. We will do that in the following.

\subsection{Entropy with island}

As we mentioned before, the generalized entropy in the presence of an island whose boundaries are located say at $a_\mp=(\mp t_a,a)$---see the configuration in the Penrose diagram of Fig.~\ref{fig:island}---receives two contributions and reads
\be
\label{eq:entrop_island_a}
S_{\rm gen} = \frac{ a^2 }{2G_N} + S_{\rm matter}\left(R\cup\text{Island}\right) \ ,
\ee
where the first term in the above corresponds to the contribution of the area of the island which is endowed with a planar and not a spherical geometry as can be seen from~(\ref{eq:metric_2D}). The exact form of the second term capturing the matter effect depends on the location of the entanglement region $R$ relative to the black hole horizon. 

By construction we are considering configurations for which the radiation is located away from the horizon ($b\gg r_h$). 
In that case, the entanglement entropy for two-dimensional massless fields living in the union of the radiation and the island regimes reads~\cite{Huerta1,Calabrese:2009ez,*Calabrese:2010he,Hashimoto:2020cas,Alishahiha:2020qza}
\be
\label{eq:entangl_entrop_island}
S_{\rm matter}\left(R\cup\text{Island}\right)= \frac{c}{3}\log\left[\frac{l(a_+,a_-)l(b_+,b_-)l(a_+,b_+)l(a_-,b_-)}{l(a_+,b_-)l(a_-,b_+)} \right] \ , 
\ee
with 
\be
l(z,z') = e^{\rho(z)}e^{\rho(z')}\sqrt{(u(z')-u(z))(v(z)-v(z'))} \ ,
\ee
the geodesic distance between $z$ and $z'$ in the geometry of~(\ref{eq:metric_2D}). 
The relation~(\ref{eq:entangl_entrop_island}) explicitly reads
\be
\begin{aligned}
S_{\rm matter}\left(R\cup\text{Island}\right) &= \frac c 6 \log \left[ 16\left(\frac{a^2-r_h^2}{r_h^2}\right)\left(\frac{b^2-r_h^2}{r_h^2}\right)\cosh^2 \frac{t_a}{r_h}\cosh^2 \frac{t_b}{r_h} \right] \\
&\qquad\qquad+\frac c 3 \log\left[\frac{\frac 1 2 \left(\sqrt{\frac{a^2-r_h^2}{b^2-r_h^2}} +\sqrt{\frac{b^2-r_h^2}{a^2-r_h^2}}\right) -\cosh\frac{(t_a-t_b)}{r_h}}{\frac 1 2 \left(\sqrt{\frac{a^2-r_h^2}{b^2-r_h^2}} +\sqrt{\frac{b^2-r_h^2}{a^2-r_h^2}}\right) +\cosh\frac{(t_a+t_b)}{r_h}}\right]\ . 
\end{aligned}
\ee
Taking into account that the island is located very close to the horizon, i.e. $a\approx r_h$,\footnote{This assumption will be justified a posteriori, see Eq.~(\ref{eq:isl_loc}).} while $b\gg r_h$ since the matter sector is far from the black hole, we find
\be
\begin{aligned}
S_{\rm matter}\left(R\cup\text{Island}\right) &\approx \frac c 6 \log \left[ 16\left(\frac{a^2-r_h^2}{r_h^2}\right)\left(\frac{b}{r_h}\right)^2\cosh^2 \frac{t_a}{r_h}\cosh^2 \frac{t_b}{r_h} \right] \\
&\qquad\qquad\qquad\qquad+\frac c 3 \log\left[\frac{1 -2\frac{\sqrt{a^2-r_h^2}}{b}\cosh\frac{(t_a-t_b)}{r_h}}{1 +2\frac{\sqrt{a^2-r_h^2}}{b}\cosh\frac{(t_a+t_b)}{r_h}}\right]\ . 
\end{aligned}
\ee
At early times $t_a,t_b\ll r_h$ so that  
\be
\begin{aligned}
S_{\rm matter}\left(R\cup\text{Island}\right) &\approx \frac c 6 \log \left[ 16\left(\frac{a^2-r_h^2}{r_h^2}\right)\left(\frac{b}{r_h}\right)^2\cosh^2 \frac{t_a}{r_h}\cosh^2\frac{t_b}{r_h} \right] \\
&\qquad\qquad\qquad\qquad\qquad-\frac {4c}{3}\frac{\sqrt{a^2-r_h^2}}{b} \cosh\frac{t_a}{r_h} \cosh \frac{t_a}{r_h} \ .
\end{aligned}
\ee
Plugging the above into~(\ref{eq:sgen}) and extremizing w.r.t. both  $t_a$ and $a$, one easily finds that there is no real saddle. This confirms our expectations that no island is present at early times and the entanglement entropy of the system is completely determined by the matter contribution only and grows with time, cf.~(\ref{eq:smatter_ein}). 

We turn to the situation at late times $t_a,t_b\gg r_h$. The fine-grained matter entropy can be approximated by 
\be
S_{\rm matter}\left(R\cup\text{Island}\right) \approx \frac {2c}{3} \log \left(\frac{b}{r_h}\right)-\frac {2c}{3} \frac{\sqrt{a^2-r_h^2}}{b}\cosh\frac{(t_a-t_b)}{r_h} \ ,
\ee
where we dropped the exponentially suppressed terms.\footnote{We could have arrived at the same expression by noting that at late times, when the left and right wedges of Fig.~\ref{fig:island} are sufficiently separated,
\be
S_{\rm matter}\left(R\cup\text{Island}\right)\approx \frac{c}{6} \log \left[(u_{b_+}-u_{a_+})^2 (v_{b_+}-v_{a_+})^2e^{2\rho(b_+)}e^{2\rho(a_+)}\right] \ ,
\ee
since it is straightforward to show that (see also~\cite{Hashimoto:2020cas})
\be
l(a_+,a_-)l(b_+,b_-)\approx l(a_+,b_-)l(a_-,b_+) \ .
\ee
}
The generalized entanglement entropy in this limit thus reads
\be
S_{\rm gen} \approx \frac{a^2}{2G_N} + \frac{2c}{3} \log \left(\frac{b}{r_h}\right)-\frac{2c}{3} \frac{\sqrt{a^2-r_h^2}}{b}\cosh\frac{(t_a-t_b)}{r_h} \ . 
\ee

Our purpose is to determine the (temporal and spatial) location of the island that extremizes the generalized entropy at late times. Let us first differentiate the above, say w.r.t. $t_a$. Requiring that this vanish, we find that the saddle point corresponds to 
\be
\label{eq:sad_point_t}
\sinh\frac{(t_a-t_b)}{r_h} = 0 \ ,
\ee
meaning that $t_a = t_b$. We now extremize $S$ w.r.t. $a$. It is straightforward to see that if 
confine ourselves to configurations subject to~(\ref{eq:sad_point_t}), we obtain immediately
\be
\label{eq:isl_loc}
a\approx r_h + \ldots \ ,
\ee
with the ellipses standing for subleading terms which, importantly, do not depend on time. 
Finally, 
\be
\label{eq:sgen_einstein}
S \approx \frac{ r_h^2}{2G_N} + \ldots \ .
\ee
which is twice the Bekenstein-Hawking entropy $S_{\rm BH}$ of the black hole.

\begin{figure}[!h]
\centering
\includegraphics[scale=.5]{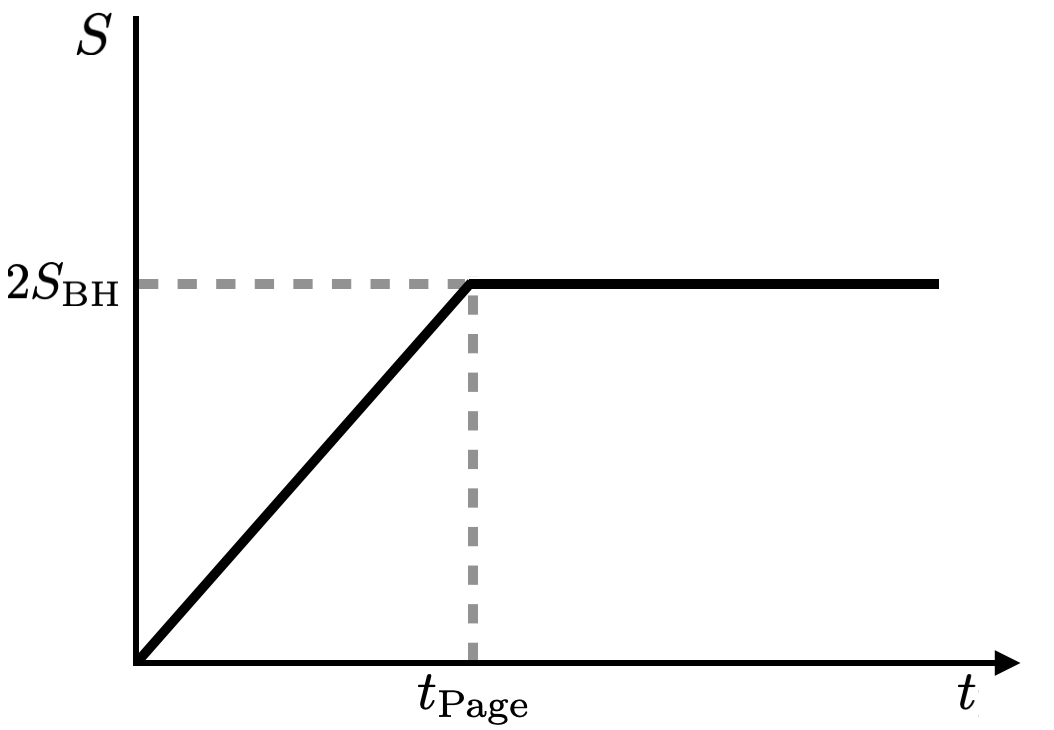}
\caption{The (qualitative) evolution of the generalized entropy of the four-dimensional linear dilaton black hole as a function of time. At approximately Page time, the contribution of the island causes the entropy to stop its monotonic growth with time and become constant.}
    \label{fig:page_curve}
\end{figure}

Let us briefly discuss the behavior of the generalized entropy, also sketched in Fig.~\ref{fig:page_curve}. At early times, it is dominated by the fine-grained entropy of the matter sector and it grows with time. Around the Page time\,\footnote{The Page time follows from equating~(\ref{eq:entr_no_island}) with $2S_{\rm BH}$.}
\be
\frac{t_{\rm Page}}{r_h} \approx \frac{3S_{\rm BH}}{2c} \ ,
\ee
one needs to take into account an island configuration located close and outside the horizon of the black hole, see~(\ref{eq:isl_loc}). Its presence is necessary to minimize the generalized entropy, which now is constant and its leading term is equal to twice the Bekenstein-Hawking entropy. 

In addition, since we were able to reproduce the correct behavior for the entropy after $t_{\rm Page}$, i.e. the fact that it  asymptotes to the Bekenstein-Hawking value, means that even though more islands may very well be present, their contributions are subdominant. 

\section{The linear dilaton black hole---String frame}
\label{sec:linear_dilaton_string}

In the previous section we worked in the Einstein frame, where the gravitational sector of the theory had its canonical, Einstein-Hilbert, form. We will now show that the details of this particular  black hole solution actually become more transparent in the string frame. 

The string-frame action for the four dimensional linear dilaton can be found by Weyl-rescaling the Einstein frame metric appearing in~(\ref{eq:dil_act_einst})
as $g_{\mu\nu} \to \tilde g_{\mu\nu} = e^\sigma g_{\mu\nu}$; it reads
\be
\label{eq:dil_act_string}
\widetilde{ I}= \frac{1}{16\pi \alpha'} \int d^4x \sqrt{\tilde g}\, e^{-\sigma}\left(\widetilde R + (\tilde \partial \sigma)^2 +4 k^2 \right) \ ,
\ee
with a tilde denoting objects defined in the string frame. 
It is a straightforward exercise to obtain the equations of motion. Varying the above w.r.t. the metric and the dilaton, we find 
\be
\label{eq:einst_string_frame}
\widetilde G_{\mu\nu} = \tilde g_{\mu\nu}\left(\widetilde \square \sigma -\frac{1}{2} (\tilde \partial \sigma)^2 +2 k^2 \right) -\widetilde\nabla_\mu \widetilde\nabla_\nu \sigma \ ,
\ee
and
\be
\widetilde\square \sigma -\frac{1}{2} (\tilde \partial \sigma)^2 +2 k^2 = -\frac{\widetilde R}{2} \ ,
\ee
respectively. Plugging the dilaton's eom into the Einstein equations, we get 
\be
\begin{aligned}
\label{eq:fixed_point}
&\widetilde R_{\mu\nu} +\widetilde\nabla_\mu \widetilde\nabla_\nu \sigma =0 \ , \\
&\widetilde R+2\widetilde\square \sigma - (\tilde \partial \sigma)^2
+4 k^2
= 0\ .
\end{aligned}
\ee

Interestingly, the above are just the conditions for conformal invariance in string theory,  upon identifying $\s =2\Phi$ as the dilaton, in the standard normalization. Indeed, the $\beta$-functions for the metric and dilaton can be calculated perturbatively in the weak string coupling expansion $\alpha'\to 0$ of the $\sigma$-model
\be
\label{eq:sigma}
I_{\rm \sigma}=\frac{1}{4\pi \alpha'}
\int d^2 \xi \sqrt{\gamma} \gamma^{ab} \tilde g_{\mu\nu}(x) \partial_a x^\mu \partial_b x^\nu +\frac{1}{4\pi}\int d^2 \xi \sqrt{\gamma} R^{(2)} \Phi(x) \ , 
\ee
and are explicitly written as 
\be
\begin{aligned}
\label{eq:beta-functions}
&\beta_{\mu\nu}^g=\alpha'\big(\widetilde R_{\mu\nu} +2\widetilde\nabla_\mu \widetilde\nabla_\nu \Phi\big) \ ,
 \\
&\beta^\Phi=-\delta c+\frac{3}{2}\alpha'\Big[  4(\tilde \partial \Phi)^2 
-4 \widetilde\square \Phi- \widetilde R\Big] \  .
\end{aligned}
\ee
Here, $\delta c$ is the central charge deficit given by
\be
\delta c=\frac{2}{3}(D_{\rm{crit}}-D_{\rm{eff}}) \ ,     
\ee
where $D_{\rm{eff}}=D$, the spacetime dimensions and $D_{\rm{crit}}=26$ for the bosonic string, whereas
$D_{\rm{eff}}=3D/2$ and $D_{\rm{crit}}=15$ for the superstring. 
Conformal invariance dictates that  $\beta_{\mu\nu}^g=\beta^\Phi=0$,  which are just 
Eqs.~(\ref{eq:fixed_point}) with
\be
\delta c= 6k^2\alpha'  \ .
\ee
It can be immediately checked that~(\ref{eq:fixed_point}) is solved by 
\bea
\label{eq:einst_fram_lin_elem}
&&\widetilde{ds}^2 =k^{-2} \left(-\left(1-\frac{r_h^2}{r^2}\right)dt^2+\frac{dr^2}{r^2\left(1-\frac{r_h^2}{r^2}\right)}+dx^2+dy^2\right)\ , \\ 
&&\sigma=-2\log (kr)-\s_0 \ , 
\eea
with $\s_0$ a constant related to the black hole mass~\cite{Witten:1991yr}. Note that, clearly, 
\be
\widetilde{ds}^2 = e^{\sigma+\sigma_0} ds^2 \ , 
\ee
with $ds^2$ is the Einstein frame line element given in Eq.~(\ref{eq:ds40}). Despite the fact that the form of the metric has changed due to the presence of the conformal factor, the causal structure of both the string- and Einstein- frame geometries is the same. This follows trivially from the fact that the Weyl rescaling relating the two frames preserves the angles. 

We may now proceed as in Sec.~\ref{sec:linear_dilaton_Einstein} and express the metric in terms of the ``light-cone'' coordinates $(u,v)$ introduced previously. It is easy to show that~(\ref{eq:einst_fram_lin_elem}) can be written as
\be
\label{eq:met_jordan_frame}
\widetilde{ds}^2 = k^{-2}\left(-\frac{du dv} {r_h^2-u v }+ dx^2+ dy^2\right) \ ,
\ee
whose $u-v$ part is nothing  more than the Witten black-hole  solution~\cite{Witten:1991yr} described by a  
2D $SL(2,R)/U(1)$ coset CFT.\footnote{The Euclidean version 
of this solution was found by Elitzur, Forge and Rabinovici~\cite{Rabino}.} In view of Eq.~(\ref{eq:fixed_point}) and its relation to string theory, this should hardly come as a surprise. 

We are now in position to repeat the computation of the entropy for the string-frame metric~(\ref{eq:met_jordan_frame}). As before, without an island, we find that the entropy 
asymptotically behaves as 
\be
\label{eq:Sw}
S=S_{\rm matter} \sim \frac{c}{3} \frac{t_b}{r_h}  \ , 
\ee
i.e. it grows linearly with time.  This result coincides with~\cite{Gautason:2020tmk,Anegawa:2020ezn} concerning the purely two-dimensional (eternal) Witten black hole.

When an island is included, an identical computation with the one carried out in the Einstein frame reveals that the generalized entropy is independent of time and its leading term is proportional to twice the Bekenstein-Hawking entropy
\be
\label{eq:Si}
S\approx\frac{k^{-2}e^{-\s}}{2\a'}\Bigg\vert_{\rm horizon} \approx\frac{r_h^2}{2 G_N}\ , 
\ee
where 
\be
G_N=\alpha'\,e^{-\s_0}\ ,
\label{GN}
\ee
and the $k^{-2}$ factor in the above is due to the fact that the scale in each of the $x$ and $y$ directions is $k^{-1}$.  The entropy formula~(\ref{eq:Si}) is identical to what we found previously~(\ref{eq:sgen_einstein}) which is of course expected, as the Bekenstein-Hawking entropy is invariant under Weyl transformations~\cite{BH2}.

\section{``Running" horizons}
\label{sec:running}

\begin{figure}[!h]
\centering
\includegraphics[scale=.5]{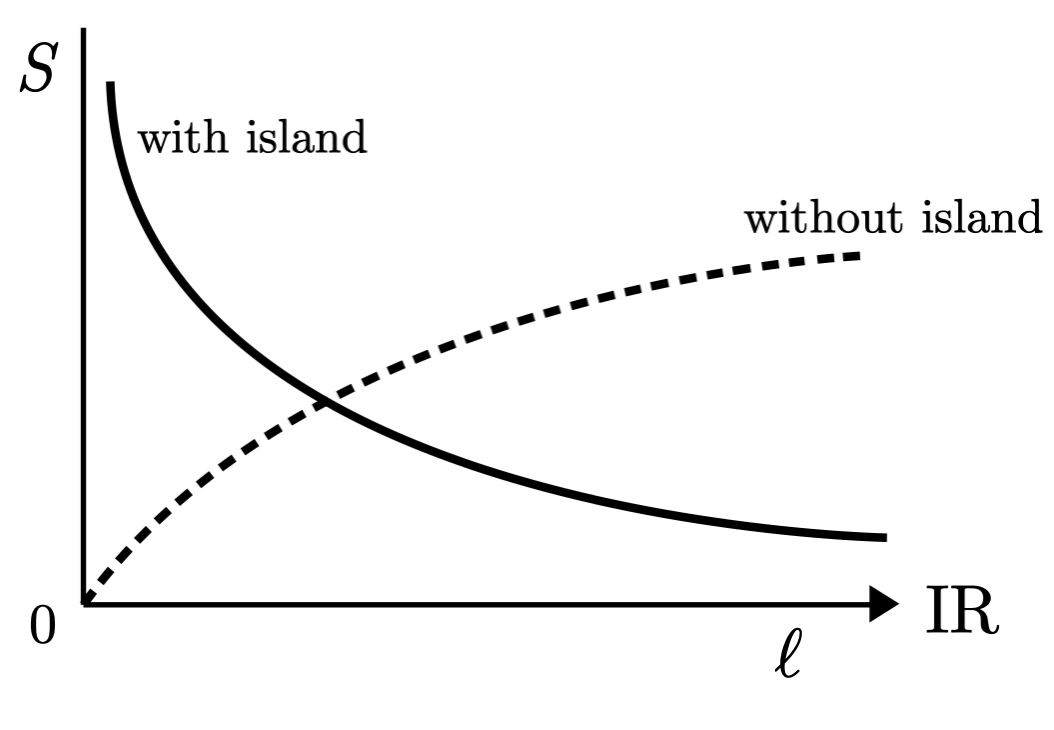}
\caption{The entropy as a function of the world-sheet length scale $\ell$. When there are no islands, the entropy is increasing towards the IR, whereas it goes to zero when an island is included.}
\label{fig:page_curve_running}
\end{figure}

The two-dimensional sigma model described by the action~(\ref{eq:sigma}) is a perturbatively renormalizable quantum field theory and the scale dependence of the generalized couplings, which are here  the target space metric $\tilde g_{\mu\nu}$ and the dilaton $\Phi$,  can be computed order by order in perturbation theory.   The 1-loop renormalization of the metric and dilaton w.r.t. the logarithm of  the world-sheet length scale $\ell$  is specified by the following  RG flow equations
\bea
&&\frac{\partial \tilde{g}_{\mu\nu}}{\partial \lambda}=-\beta_{\mu\nu}^g=-\alpha'\big(\widetilde R_{\mu\nu} +2\widetilde\nabla_\mu \widetilde\nabla_\nu \Phi\big) \ , \label{eq:RF1}
\\
&&\frac{\partial \Phi}{\partial  \lambda}=-\beta^\Phi=\delta c-\frac{3}{2}\alpha'\Big[  4(\tilde \partial \Phi)^2 
-4 \widetilde\square \Phi- \widetilde R\Big] \ ,\label{eq:RF2}
\eea
where $\lambda=\log \ell/\ell_0$ and $\ell_0$ a reference scale.
Then, the solution~(\ref{eq:einst_fram_lin_elem}) is a fixed point of the RG flow equations\,\footnote{  Eq.~(\ref{eq:RF1}) is known as Ricci flow in the mathematical literature and its fixed points as solitons
\cite{Chow,Bakas,KLL1}. } 
\be
    \beta^g_{\mu\nu}=\beta^\Phi=0 \ . 
\ee
However, one may look for more general solutions to~(\ref{eq:RF1}) and~(\ref{eq:RF2}), for which the beta functions need not vanish. 
It is not difficult to verify that one such class of solutions reads 
\bea
&&ds^2=-k_s^{-2}\frac{dudv}{r_0^2e^{2a\lambda}-uv} +dx^2+dy^2 \ , \label{eq:Rm1}\\
&&\Phi=-\frac{3}{5}\log\big[k_s^2(r_0^2e^{2a\lambda}-uv)\big]+\Phi_0 \ ,\label{eq:Rm2}
\eea
where $x,y$ are two free bosons and 
\be
k_s^2\alpha'=-5a \ ,
~~~a=-\frac{5\delta c}{216} < 0 \ . 
\ee
Comparing~(\ref{eq:Rm1}) with~(\ref{eq:met_jordan_frame}), we see that the horizon $r_h$ of this geometry is identified with\,\footnote{As in the (string-frame) linear dilaton black hole, in the $(u,v)$ coordinate system the singularity is located at $uv= r_h^2$, while the horizon at $u v =0$.}
\be
r_h=r_0e^{a\lambda}=r_0 \left(\frac{\ell}{\ell_0}\right)^{-\frac{5\delta c}{216}} \ , 
\ee
with $r_0$ the value of $r$ at the worldsheet length scale $\ell_0$ (i.e. $\lambda=0)$. 
Therefore, $r_h$ shrinks as we approach the IR
\be
r_h\to 0 \ ,~~~\text{as}~~~\lambda\to \infty~~(\ell\to \infty) \ . 
\ee
Therefore, without an island, the entropy at any fixed time $t_b$ is increasing towards the IR 
as it can be seen from Eq.~(\ref{eq:Sw}). Indeed, in this case we find that 
\be
S\approx \frac{c}{3}\frac{t_b}{r_0}e^{-a\lambda} \ ,
\ee
which blows up asymptotically at the IR.
On the other hand, when an island is included, 
the entropy scales as 
\be
S\approx \frac{ r_h^2}{2G_N}=\frac{ r_0^2}{2G_N} e^{2a\lambda} \ ,
\ee
and is now decreasing towards the IR. 
This behavior of the running entropy is depicted in Fig.~\ref{fig:page_curve_running}.
Note that the fixed point of the RG running corresponds to the deep IR limit $\lambda \to \infty$, where the metric describes four free
bosons and a linear dilaton and therefore the entropy should vanish asymptotically there, something which is also consistent with the c-theorem~\cite{Huerta1,Huerta2,Dimitris}. This behavior is restored only when an island is included.

Configurations such as the ones described by Eqs.~(\ref{eq:Rm1}) and~(\ref{eq:Rm2}) may be thought of as being ``off-shell,'' in the sense that they do not  solve~(\ref{eq:fixed_point}). Thus,  they are not saddles of the string effective action~(\ref{eq:dil_act_string}), so it is reasonable for one to wonder whether the island prescription is applicable or even meaningful at all in this situation. Let us now understand why the answer is positive. 

It can be easily verified that configurations of the form 
 \bea
&&ds^2=-k_s^{-2}\frac{dudv}{r_0^2e^{2a\lambda}-uv} +dx^2+dy^2 \ , \label{gen1}\\
&&\Phi_\g=-\frac{1}{\g}\log\big[k_s^2(r_0^2e^{2a\lambda}-uv)\big]+\Phi_0 \ ,\label{gen2}
\eea
with $\g$ a constant, satisfy 
\be
\begin{aligned}
\label{eq:eom_phig}
&\widetilde R_{\mu\nu}+\g \widetilde\nabla_\mu\widetilde\nabla_\nu \Phi_\g=0 
 , \\
&\widetilde R+2\g \widetilde\Box \Phi_\g-\g^2 (\tilde\partial\Phi_\g)^2+4 k_s^2=0 \ ,
\end{aligned}
\ee
which, in turn means that~\emph{they are saddles of}
\be
\label{eq:act_phig}
\widetilde I_\g=\frac{1}{16\pi \alpha'} \int d^4 x \sqrt{\tilde g}e^{-\g \Phi_\g}\Big(\widetilde R+\g^2 (\tilde \partial\Phi_\g)^2 +4k_s^2 \Big)  \ .
\ee 
This is of course the case for arbitrary values of the parameters that characterize the solution. In particular, the case $\gamma=2$ and $a=0$ corresponds to the string effective action, while for $\g=5/3$ and $a< 0$, we get the running-horizons geometries~(\ref{eq:Rm1}) and~(\ref{eq:Rm2}). In other words, one can think of these black holes either as a particular solution of the RG flow equations~(\ref{eq:RF1}) and~(\ref{eq:RF2}), or as saddles of the action~(\ref{eq:act_phig}).\footnote{
Note that using $\Phi=\Phi_{\frac{5}{3}}$ in~(\ref{eq:sigma}) gives rise to a 2D conformal anomaly and this is the reason for the non-vanishing of the $\beta$-functions~(\ref{eq:beta-functions}) and the corresponding running.} This is the reason that the island prescription also works here.
 
 A final comment concerns the fixed points of two-dimensional world-sheet RG flows, which may very well exist both in the UV and IR domains. As we have seen above, there is an IR fixed point describing four free bosons, whereas the UV limit is singular.  However, generally,  RG flows tend to develop singularities~\cite{Chow}; these  can be past (UV) or future (IR) ones and are evolved in the so-called ancient and immortal solutions, respectively. In our case, it so happens that there is a UV singularity, something in accordance with the Ricci-flow theory.

\section{A charged dilaton black hole and its entanglement entropy}
\label{sec:charged}

Let us now include a $U(1)$ gauge field in our discussion. The action in this case is given by 
\be 
I=\frac{1}{16\pi G_N}\int d^4x \sqrt g\left(R-\f{1}{2}(\p \s)^2+4k^2e^\s-\f{1}{4}e^{\gamma \s}\,F^2\right) \ ,
\ee
where $\gamma$ is a constant and $F^2\equiv F_{\mu\nu}F^{\mu\nu}$, with  $F_{\mu\nu}=\p_{\mu}A_{\nu}-\p_{\nu}A_{\mu}$, the $U(1)$ field strength.
The equations of motion can be easily obtained by varying the above w.r.t. $g_{\mu\nu}$, $A_\mu$ and $\s$:
\bea
& R_{\mu\nu}-\frac{1}{2}g_{\mu\nu}R=\frac{1}{2}(\p_\mu \s)(\p_\nu \s)-\frac{1}{2}g_{\mu\nu}\Bigg(\frac{1}{2}(\p \s)^2-4k^2e^\s\Bigg)+\frac{1}{2}e^{\gamma\s}\Bigg(F_{\mu\alpha}F_\nu^{~\alpha}
-\f{1}{4}g_{\mu\nu}F^2\Bigg) \ ,&\\
& \nabla_\mu\Big(e^{\gamma \s}F^{\mu\nu}\Big) = 0 \ ,& \label{eq:Max3eq}\\
& \Box \s=-4k^2e^\s+\f{1}{4}\gamma e^{\gamma \s} F^2 \ .&\label{eq:S3eq}
\eea
Let us consider the following ansatz for the metric 
\be
\label{eq:ds3}
ds^2=-b(r)dt^2+a(r) dr^2+r^{2}\Big(dx^2+dy^2\Big)
 \ ,
\ee
where $b(r) =r^2/a(r) $. Like in the uncharged solution, we will insist on the dilaton being the following logarithm of the radial coordinate $r$
\be
\label{eq:dilaton_charged_BH}
\s = -2 \log (k r) \ .
\ee
On top of~(\ref{eq:ds3}) and~(\ref{eq:dilaton_charged_BH}), we can immediately solve Maxwell equations~(\ref{eq:Max3eq}), by requiring that 
\be
A_t = - \frac{C}{2\g} (k r)^{2\gamma} \ ,~~~A_r=A_x=A_y= 0 \ ,
\ee
or in other words
\be 
\label{eq:Max3sol}
F_{\mu\nu}=-
\f{2C (k r)^{2\gamma}}{r^3}g_{r[\mu}g_{\nu]t} \ ,
\ee
with $C$ a (nonzero) constant and the square brackets stand for antisymmetrization of the corresponding indexes. 

The requirement that the remaining equations of motion be simultaneously satisfied dictates that
\be
a(r) = \frac{1}{1-\frac{r_h^2}{r^2}+\frac{Q^2}{4 r^4}}\ ,~~~b(r)=r^2\left(1-\frac{r_h^2}{r^2} + \frac{Q^2}{4 r^4}\right)\ ,~~~\g = -1 \ ,
\ee
where we introduced the ``charge'' $Q=C/\sqrt{2}k$. 
Notice that for $Q\to0$ (or equivalently $C\to 0$), the above  boils down to the uncharged solution we found in the main text, cf.~(\ref{eq:s1}).
Consequently, the charged linear dilaton black hole line element explicitly reads
\be
\label{eq:charged_simple}
ds^2 = -r^2\left(1-\f{r_h^2}{r^2}+\f{Q^2}{4r^4}\right)dt^2 + \frac{dr^2}{1-\f{r_h^2}{r^2}+\f{Q^2}{4r^4}}+r^2 (dx^2+dy^2) \ .  \ee
The black hole admits the following outer ($r_+$) and inner ($r_-$) horizons 
\be
r_{\pm}=\frac{r_h}{\sqrt 2}\left[1\pm \left(1-\frac{Q^2}{r_h^4}\right)^{1/2}\right]^{1/2} \ .
\ee
It is clear from the above that the two horizons coalesce when 
\be
\label{eq:charge_value}
Q=r_h^2 \ ,
\ee
meaning that for this value of the charge, the black hole becomes extremal with line element 
\be
\label{eq:charged_simple_extremal}
ds^2 = -r^2\left(1-\f{r_h^2}{2r^2}\right)^2 dt^2 + \frac{dr^2}{\left(1-\f{r_h^2}{2r^2}\right)^2}+r^2 (dx^2+dy^2) \ ,
\ee
as it can be straightforwardly seen from~(\ref{eq:charged_simple}) and~(\ref{eq:charge_value}). 

In the following considerations we confine ourselves to the simplest, extremal configuration~(\ref{eq:charged_simple_extremal}), leaving the investigation of the more general case for future work.  

To find the entanglement entropy of this black hole we have to  practically repeat what we did in the uncharged case.   The starting point is to rescale $t\to r_h t$ and introduce  Kruskal-type coordinates ($u$, $v$), such that the line element of the extremal solution is brought to the form~(\ref{eq:metric_2D}). To find the tortoise coordinate, we consider null geodesics 
\be
dt = r_h\frac{dr}{r\left(1-\f{r_h^2}{2r^2}\right)^2} \ ,
\ee
translating into 
\be
\label{eq:tort_charged}
r^* 
= \frac {r_h}{2} \left[ \log\left(\frac{2r^2}{r_h^2}-1\right) -  \frac{\frac{r_h^2}{2r^2}}{1-\frac{r_h^2}{2r^2}}  \right] \ .
\ee
In terms of the above, the line element becomes 
\be
ds^2  = -\left(\frac{r}{r_h}\right)^2\left(1-\f{r_h^2}{2r^2}\right)^2 \left(dt^2 -dr^{*2}\right)+r^2\left(dx^2+dy^2\right) \ ,
\ee
with 
\be
r\equiv r(r^*) = \frac{r_h}{\sqrt{2}}\left[1+ \frac{1}{\mathcal W\left(e^{-\frac{2r^*}{r_h}}\right)}\right]^{1/2}\ ,
\ee
where $\mathcal W(x)$ is the Lambert W-function. 
 
With the same null coordinates $u$ and $v$ introduced previously, see~(\ref{eq:uvstar}) and~(\ref{eq:uv}), we find that the charged metric is expressed in the desirable form
\be
\label{eq:uv_charged}
ds^2 = -e^{2\rho} du dv +r^2\left(dx^2+dy^2\right) \ ,
\ee
where the ``conformal factor'' in terms of the global coordinate $r$ is in this case given by
\be
\rho =  \log\left[\frac{r}{r_h}\left(1-\frac{r_h^2}{2r^2}\right)\right]-\frac{r^*}{r_h} \ ,
\ee
and $r^*$ can be read from~(\ref{eq:tort_charged}). 
Using the above we can compute the entanglement entropy per unit area without and with an island contribution,  using Eqs.~(\ref{eq:entr_formula_noisl})  and~(\ref{eq:entangl_entrop_island}), respectively. In the absence of an island, we obtain 
\be
S = \frac c 3 \log\left[ \frac{ (2b^2-r_h^2)}{b r_h} \cosh \frac{t_b}{r_h}\right]\approx \frac {c} {3} \log \left[\frac{2b}{r_h}\cosh \frac{t_b}{r_h}\right] \ ,
\ee
which for $b\gg r_h$ coincides with  what we got for the uncharged black hole. Including an island very close but slightly outside the black hole's horizon at $a=r_h/\sqrt{2}$, we find that, exactly as before, the value of the entropy at late times asymptotes to twice the Bekenstein-Hawking one for the extremal configuration.

\section{Discussion}
\label{sec:conclusion}

In this paper we derived a novel black hole geometry in the four dimensional linear dilaton theory. Interestingly, its causal structure is the same as in the well-known Schwarzschild solution, nevertheless, it does not asymptote to the Minkowski metric and has a planar horizon. 
For this black hole, we employed the ``island prescription'' to compute the entanglement entropy of the Hawking radiation. The fact that the black hole is endowed with a planar horizon makes it necessary to work in terms of quantities defined per unit area. This has far-reaching consequences, since it enabled us to use well-known results valid in two dimensions concerning the system's entropy. We demonstrated that in order for the entropy to follow a Page curve, a contribution stemming from an island configuration should be included around the Page time. This result is in agreement with the expectation that the range of applicability of the aforementioned technique covers non-standard black hole geometries. 
 
We showed that when Weyl-transformed in the string frame, the  radial-temporal piece of our solution is nothing more than the well-known Witten black hole geometry. That this is the case could be guessed by inspection of the equations of motion in this frame: they correspond to the four-dimensional generalization of the RG equation's fixed points of the (two dimensional) world-sheet sigma model. An analogous computation of the entanglement entropy yields the same behavior, as expected.

The theory we studied descends from the stringy sigma model  with zero beta functions for the metric and dilaton. Moving away from the fixed point, we constructed a broad class of black hole geometries that satisfy the RG equations. Their corresponding entanglement entropy depends non trivially on the worldsheet length scale. As we flow towards the IR, the entropy before Page's time increases, but when the island kicks in, it decays asymptotically. 

Finally, when  a $U(1)$ gauge sector is coupled to the linear dilaton model, new solutions to the equations of motion become accessible to the system. One such geometry is a charged extremal black hole, for which the expected late-time behavior  of its entanglement entropy is a result of an island contribution.

\section*{Acknowledgments}

We are grateful to Gia Dvali for correspondence and important discussions and Dimitrios Giataganas for discussions.

\appendix

\section{ The asymptotic linear dilaton background}
\label{app:background}

Yet another (and probably the simplest) solution to the field equations~(\ref{eq:einst_eqs_cov}) and~(\ref{eq:dil_eom_cov}) is
\bea
&&d s^2=-r^2 dt^2 + d r^2+r^2\left(d x^2+d y^2\right) \ , \label{eq:ds4} \\
&&\sigma=-2\, \log (k r) \ . \label{eq:s}
 \eea
It is obvious that the above metric is the asymptotic, $r\to\infty$, limit of the black hole geometry~(\ref{eq:ds40}).

Actually,~(\ref{eq:ds4}) is  the induced metric on the 5D Lorentzian cone
\be
\label{eq:cone}
-X_0^2+X_1^2+X_2^2-2X_3^2+2 X_3 X_4=0 \ , 
\ee
embedded in $M^{3,2}$ with line element 
\be
d s_5^2= -d X_0^2+d X_1^2+d X_2^2 -d X_3^2+d X_4^2 \ .
\ee
It can be immediately verified that the  parametrization 
\be
X_0=rt,~~~X_1=rx,~~~X_2=ry,~~~X_3=\frac{r}{2}\Big(-t^2+x^2+y^2\Big),~~~X_4=X_3-r,
\ee
leads to the induced metric~(\ref{eq:ds4}) on the cone~(\ref{eq:cone}). 
 
The coordinate transformation $r=k^{-1} e^z$, brings~(\ref{eq:ds4}) in the more familiar form
\bea
&&d s^2=k^{-2}e^{2z}\Big(-dt^2+d x^2+d y^2+d z^2\Big) \ , \label{eq:conmetric}\\
&&\sigma =-2z \ ,
\eea 
which is the linear dilaton sting background in the Einstein frame. It has also appeared as the ``continuous clockwork
geometry'' in~\cite{Giudice1,Giudice2,KR1,KR2}.

Note that the metric is singular at $r=0$ since curvature invariants diverge there; for example, a straightforward computation reveals that
\be
R= -\frac{6}{r^2} \ ,~~~  R_{\mu\nu}^2=\frac{12}{r^4} \ ,~\ldots
\ee
This singularity is a naked one, so it should be dressed. In the clockwork case 
it is completely cut out of the spacetime by the introduction of end-of-the world branes at $r=1, (z=0)$ and $r=e^{z_0}$.   
However, there is another  possibility. Namely, to hide the singularity behind a horizon at, say,  $r_h$. This situation corresponds precisely to the black hole solution we derived in the main text, see Sec.~\ref{sec:linear_dilaton_Einstein}, Eqs.~(\ref{eq:ds40}) and~(\ref{eq:s1}); for the convenience of the reader, we present it here as well:
\bea
&&d s^2=-r^2\left(1-\frac{r_h^2}{r^2}\right)d t^2+\frac{d r^2}{1-\frac{r_h^2}{r^2}}+r^2\left(
d x^2+d y^2\right) \ ,\label{ds40}\\
&&\sigma=-2\, \log (k r)\ .  
\eea
We observe that now the singularity ($r=0$) is located behind the horizon at $r=r_h$, which~\emph{is flat and not spherical}. 
The geometry endowed with the metric~(\ref{ds40}) is what we identify as the ``linear dilaton black hole'' in this work.

\bibliographystyle{utphys}
\bibliography{BH_Linear_Dilaton.bib}

\providecommand{\href}[2]{#2}\begingroup\raggedright\begin{thebibliography}{10}

\bibitem{Dvali:2011aa}
G.~Dvali and C.~Gomez, ``{Black Hole's Quantum N-Portrait},''
  \href{http://dx.doi.org/10.1002/prop.201300001}{{\em Fortsch. Phys.}
  {\bfseries 61} (2013) 742--767},
  \href{http://arxiv.org/abs/1112.3359}{{\ttfamily arXiv:1112.3359 [hep-th]}}.

\bibitem{Dvali:2012rt}
G.~Dvali and C.~Gomez, ``{Black Hole's 1/N Hair},''
  \href{http://dx.doi.org/10.1016/j.physletb.2013.01.020}{{\em Phys. Lett. B}
  {\bfseries 719} (2013) 419--423},
  \href{http://arxiv.org/abs/1203.6575}{{\ttfamily arXiv:1203.6575 [hep-th]}}.

\bibitem{Dvali:2012wq}
G.~Dvali and C.~Gomez, ``{Black Hole Macro-Quantumness},''
  \href{http://arxiv.org/abs/1212.0765}{{\ttfamily arXiv:1212.0765 [hep-th]}}.

\bibitem{Dvali:2015aja}
G.~Dvali, ``{Non-Thermal Corrections to Hawking Radiation Versus the
  Information Paradox},'' \href{http://dx.doi.org/10.1002/prop.201500096}{{\em
  Fortsch. Phys.} {\bfseries 64} (2016) 106--108},
  \href{http://arxiv.org/abs/1509.04645}{{\ttfamily arXiv:1509.04645
  [hep-th]}}.

\bibitem{Almheiri:2020cfm}
A.~Almheiri, T.~Hartman, J.~Maldacena, E.~Shaghoulian, and A.~Tajdini, ``{The
  entropy of Hawking radiation},''
  \href{http://arxiv.org/abs/2006.06872}{{\ttfamily arXiv:2006.06872
  [hep-th]}}.

\bibitem{Penington:2019npb}
G.~Penington, ``{Entanglement Wedge Reconstruction and the Information
  Paradox},'' \href{http://dx.doi.org/10.1007/JHEP09(2020)002}{{\em JHEP}
  {\bfseries 09} (2020) 002}, \href{http://arxiv.org/abs/1905.08255}{{\ttfamily
  arXiv:1905.08255 [hep-th]}}.

\bibitem{Almheiri:2019psf}
A.~Almheiri, N.~Engelhardt, D.~Marolf, and H.~Maxfield, ``{The entropy of bulk
  quantum fields and the entanglement wedge of an evaporating black hole},''
  \href{http://dx.doi.org/10.1007/JHEP12(2019)063}{{\em JHEP} {\bfseries 12}
  (2019) 063}, \href{http://arxiv.org/abs/1905.08762}{{\ttfamily
  arXiv:1905.08762 [hep-th]}}.

\bibitem{Almheiri:2019hni}
A.~Almheiri, R.~Mahajan, J.~Maldacena, and Y.~Zhao, ``{The Page curve of
  Hawking radiation from semiclassical geometry},''
  \href{http://dx.doi.org/10.1007/JHEP03(2020)149}{{\em JHEP} {\bfseries 03}
  (2020) 149}, \href{http://arxiv.org/abs/1908.10996}{{\ttfamily
  arXiv:1908.10996 [hep-th]}}.

\bibitem{Almheiri:2019yqk}
A.~Almheiri, R.~Mahajan, and J.~Maldacena, ``{Islands outside the horizon},''
  \href{http://arxiv.org/abs/1910.11077}{{\ttfamily arXiv:1910.11077
  [hep-th]}}.

\bibitem{Chen:2019uhq}
H.~Z. Chen, Z.~Fisher, J.~Hernandez, R.~C. Myers, and S.-M. Ruan,
  ``{Information Flow in Black Hole Evaporation},''
  \href{http://dx.doi.org/10.1007/JHEP03(2020)152}{{\em JHEP} {\bfseries 03}
  (2020) 152}, \href{http://arxiv.org/abs/1911.03402}{{\ttfamily
  arXiv:1911.03402 [hep-th]}}.

\bibitem{Almheiri:2019psy}
A.~Almheiri, R.~Mahajan, and J.~E. Santos, ``{Entanglement islands in higher
  dimensions},'' \href{http://dx.doi.org/10.21468/SciPostPhys.9.1.001}{{\em
  SciPost Phys.} {\bfseries 9} no.~1, (2020) 001},
  \href{http://arxiv.org/abs/1911.09666}{{\ttfamily arXiv:1911.09666
  [hep-th]}}.

\bibitem{Penington:2019kki}
G.~Penington, S.~H. Shenker, D.~Stanford, and Z.~Yang, ``{Replica wormholes and
  the black hole interior},'' \href{http://arxiv.org/abs/1911.11977}{{\ttfamily
  arXiv:1911.11977 [hep-th]}}.

\bibitem{Almheiri:2019qdq}
A.~Almheiri, T.~Hartman, J.~Maldacena, E.~Shaghoulian, and A.~Tajdini,
  ``{Replica Wormholes and the Entropy of Hawking Radiation},''
  \href{http://dx.doi.org/10.1007/JHEP05(2020)013}{{\em JHEP} {\bfseries 05}
  (2020) 013}, \href{http://arxiv.org/abs/1911.12333}{{\ttfamily
  arXiv:1911.12333 [hep-th]}}.

\bibitem{Chen:2019iro}
Y.~Chen, ``{Pulling Out the Island with Modular Flow},''
  \href{http://dx.doi.org/10.1007/JHEP03(2020)033}{{\em JHEP} {\bfseries 03}
  (2020) 033}, \href{http://arxiv.org/abs/1912.02210}{{\ttfamily
  arXiv:1912.02210 [hep-th]}}.

\bibitem{Bhattacharya:2020ymw}
A.~Bhattacharya, ``{Multipartite purification, multiboundary wormholes, and
  islands in $AdS_3/CFT_2$},''
  \href{http://dx.doi.org/10.1103/PhysRevD.102.046013}{{\em Phys. Rev. D}
  {\bfseries 102} no.~4, (2020) 046013},
  \href{http://arxiv.org/abs/2003.11870}{{\ttfamily arXiv:2003.11870
  [hep-th]}}.

\bibitem{Gautason:2020tmk}
F.~F. Gautason, L.~Schneiderbauer, W.~Sybesma, and L.~Thorlacius, ``{Page Curve
  for an Evaporating Black Hole},''
  \href{http://dx.doi.org/10.1007/JHEP05(2020)091}{{\em JHEP} {\bfseries 05}
  (2020) 091}, \href{http://arxiv.org/abs/2004.00598}{{\ttfamily
  arXiv:2004.00598 [hep-th]}}.

\bibitem{Anegawa:2020ezn}
T.~Anegawa and N.~Iizuka, ``{Notes on islands in asymptotically flat 2d dilaton
  black holes},'' \href{http://dx.doi.org/10.1007/JHEP07(2020)036}{{\em JHEP}
  {\bfseries 07} (2020) 036}, \href{http://arxiv.org/abs/2004.01601}{{\ttfamily
  arXiv:2004.01601 [hep-th]}}.

\bibitem{Hashimoto:2020cas}
K.~Hashimoto, N.~Iizuka, and Y.~Matsuo, ``{Islands in Schwarzschild black
  holes},'' \href{http://dx.doi.org/10.1007/JHEP06(2020)085}{{\em JHEP}
  {\bfseries 06} (2020) 085}, \href{http://arxiv.org/abs/2004.05863}{{\ttfamily
  arXiv:2004.05863 [hep-th]}}.

\bibitem{Hartman:2020swn}
T.~Hartman, E.~Shaghoulian, and A.~Strominger, ``{Islands in Asymptotically
  Flat 2D Gravity},'' \href{http://dx.doi.org/10.1007/JHEP07(2020)022}{{\em
  JHEP} {\bfseries 07} (2020) 022},
  \href{http://arxiv.org/abs/2004.13857}{{\ttfamily arXiv:2004.13857
  [hep-th]}}.

\bibitem{Hollowood:2020cou}
T.~J. Hollowood and S.~P. Kumar, ``{Islands and Page Curves for Evaporating
  Black Holes in JT Gravity},''
  \href{http://dx.doi.org/10.1007/JHEP08(2020)094}{{\em JHEP} {\bfseries 08}
  (2020) 094}, \href{http://arxiv.org/abs/2004.14944}{{\ttfamily
  arXiv:2004.14944 [hep-th]}}.

\bibitem{Krishnan:2020oun}
C.~Krishnan, V.~Patil, and J.~Pereira, ``{Page Curve and the Information
  Paradox in Flat Space},'' \href{http://arxiv.org/abs/2005.02993}{{\ttfamily
  arXiv:2005.02993 [hep-th]}}.

\bibitem{Alishahiha:2020qza}
M.~Alishahiha, A.~Faraji~Astaneh, and A.~Naseh, ``{Island in the Presence of
  Higher Derivative Terms},'' \href{http://arxiv.org/abs/2005.08715}{{\ttfamily
  arXiv:2005.08715 [hep-th]}}.

\bibitem{Banks:2020zrt}
T.~Banks, ``{Microscopic Models of Linear Dilaton Gravity and Their
  Semi-classical Approximations},''
  \href{http://arxiv.org/abs/2005.09479}{{\ttfamily arXiv:2005.09479
  [hep-th]}}.

\bibitem{Geng:2020qvw}
H.~Geng and A.~Karch, ``{Massive islands},''
  \href{http://dx.doi.org/10.1007/JHEP09(2020)121}{{\em JHEP} {\bfseries 09}
  (2020) 121}, \href{http://arxiv.org/abs/2006.02438}{{\ttfamily
  arXiv:2006.02438 [hep-th]}}.

\bibitem{Chen:2020uac}
H.~Z. Chen, R.~C. Myers, D.~Neuenfeld, I.~A. Reyes, and J.~Sandor, ``{Quantum
  Extremal Islands Made Easy, Part I: Entanglement on the Brane},''
  \href{http://dx.doi.org/10.1007/JHEP10(2020)166}{{\em JHEP} {\bfseries 10}
  (2020) 166}, \href{http://arxiv.org/abs/2006.04851}{{\ttfamily
  arXiv:2006.04851 [hep-th]}}.

\bibitem{Chandrasekaran:2020qtn}
V.~Chandrasekaran, M.~Miyaji, and P.~Rath, ``{Including contributions from
  entanglement islands to the reflected entropy},''
  \href{http://dx.doi.org/10.1103/PhysRevD.102.086009}{{\em Phys. Rev. D}
  {\bfseries 102} no.~8, (2020) 086009},
  \href{http://arxiv.org/abs/2006.10754}{{\ttfamily arXiv:2006.10754
  [hep-th]}}.

\bibitem{Li:2020ceg}
T.~Li, J.~Chu, and Y.~Zhou, ``{Reflected Entropy for an Evaporating Black
  Hole},'' \href{http://dx.doi.org/10.1007/JHEP11(2020)155}{{\em JHEP}
  {\bfseries 11} (2020) 155}, \href{http://arxiv.org/abs/2006.10846}{{\ttfamily
  arXiv:2006.10846 [hep-th]}}.

\bibitem{Bak:2020enw}
D.~Bak, C.~Kim, S.-H. Yi, and J.~Yoon, ``{Unitarity of Entanglement and Islands
  in Two-Sided Janus Black Holes},''
  \href{http://arxiv.org/abs/2006.11717}{{\ttfamily arXiv:2006.11717
  [hep-th]}}.

\bibitem{Bousso:2020kmy}
R.~Bousso and E.~Wildenhain, ``{Gravity/ensemble duality},''
  \href{http://dx.doi.org/10.1103/PhysRevD.102.066005}{{\em Phys. Rev. D}
  {\bfseries 102} no.~6, (2020) 066005},
  \href{http://arxiv.org/abs/2006.16289}{{\ttfamily arXiv:2006.16289
  [hep-th]}}.

\bibitem{Hollowood:2020kvk}
T.~J. Hollowood, S.~Prem~Kumar, and A.~Legramandi, ``{Hawking radiation
  correlations of evaporating black holes in JT gravity},''
  \href{http://dx.doi.org/10.1088/1751-8121/abbc51}{{\em J. Phys. A} {\bfseries
  53} no.~47, (2020) 475401}, \href{http://arxiv.org/abs/2007.04877}{{\ttfamily
  arXiv:2007.04877 [hep-th]}}.

\bibitem{Krishnan:2020fer}
C.~Krishnan, ``{Critical Islands},''
  \href{http://arxiv.org/abs/2007.06551}{{\ttfamily arXiv:2007.06551
  [hep-th]}}.

\bibitem{Engelhardt:2020qpv}
N.~Engelhardt, S.~Fischetti, and A.~Maloney, ``{Free Energy from Replica
  Wormholes},'' \href{http://arxiv.org/abs/2007.07444}{{\ttfamily
  arXiv:2007.07444 [hep-th]}}.

\bibitem{Karlsson:2020uga}
A.~Karlsson, ``{Replica wormhole and island incompatibility with monogamy of
  entanglement},'' \href{http://arxiv.org/abs/2007.10523}{{\ttfamily
  arXiv:2007.10523 [hep-th]}}.

\bibitem{Gomez:2020yef}
C.~Gomez, ``{The information of the information paradox: On the quantum
  information meaning of Page curve},''
  \href{http://arxiv.org/abs/2007.11508}{{\ttfamily arXiv:2007.11508
  [hep-th]}}.

\bibitem{Chen:2020jvn}
H.~Z. Chen, Z.~Fisher, J.~Hernandez, R.~C. Myers, and S.-M. Ruan,
  ``{Evaporating Black Holes Coupled to a Thermal Bath},''
  \href{http://arxiv.org/abs/2007.11658}{{\ttfamily arXiv:2007.11658
  [hep-th]}}.

\bibitem{Hartman:2020khs}
T.~Hartman, Y.~Jiang, and E.~Shaghoulian, ``{Islands in cosmology},''
  \href{http://dx.doi.org/10.1007/JHEP11(2020)111}{{\em JHEP} {\bfseries 11}
  (2020) 111}, \href{http://arxiv.org/abs/2008.01022}{{\ttfamily
  arXiv:2008.01022 [hep-th]}}.

\bibitem{Balasubramanian:2020coy}
V.~Balasubramanian, A.~Kar, and T.~Ugajin, ``{Entanglement between two disjoint
  universes},'' \href{http://arxiv.org/abs/2008.05274}{{\ttfamily
  arXiv:2008.05274 [hep-th]}}.

\bibitem{Balasubramanian:2020xqf}
V.~Balasubramanian, A.~Kar, and T.~Ugajin, ``{Islands in de Sitter space},''
  \href{http://arxiv.org/abs/2008.05275}{{\ttfamily arXiv:2008.05275
  [hep-th]}}.

\bibitem{Sybesma:2020fxg}
W.~Sybesma, ``{Pure de Sitter space and the island moving back in time},''
  \href{http://arxiv.org/abs/2008.07994}{{\ttfamily arXiv:2008.07994
  [hep-th]}}.

\bibitem{Chen:2020hmv}
H.~Z. Chen, R.~C. Myers, D.~Neuenfeld, I.~A. Reyes, and J.~Sandor, ``{Quantum
  Extremal Islands Made Easy, Part II: Black Holes on the Brane},''
  \href{http://dx.doi.org/10.1007/JHEP12(2020)025}{{\em JHEP} {\bfseries 12}
  (2020) 025}, \href{http://arxiv.org/abs/2010.00018}{{\ttfamily
  arXiv:2010.00018 [hep-th]}}.

\bibitem{Ling:2020laa}
Y.~Ling, Y.~Liu, and Z.-Y. Xian, ``{Island in Charged Black Holes},''
  \href{http://arxiv.org/abs/2010.00037}{{\ttfamily arXiv:2010.00037
  [hep-th]}}.

\bibitem{Bhattacharya:2020uun}
A.~Bhattacharya, A.~Chanda, S.~Maulik, C.~Northe, and S.~Roy, ``{Topological
  shadows and complexity of islands in multiboundary wormholes},''
  \href{http://dx.doi.org/10.1007/JHEP02(2021)152}{{\em JHEP} {\bfseries 02}
  (2021) 152}, \href{http://arxiv.org/abs/2010.04134}{{\ttfamily
  arXiv:2010.04134 [hep-th]}}.

\bibitem{Marolf:2020rpm}
D.~Marolf and H.~Maxfield, ``{Observations of Hawking radiation: the Page curve
  and baby universes},'' \href{http://arxiv.org/abs/2010.06602}{{\ttfamily
  arXiv:2010.06602 [hep-th]}}.

\bibitem{Hernandez:2020nem}
J.~Hernandez, R.~C. Myers, and S.-M. Ruan, ``{Quantum Extremal Islands Made
  Easy, PartIII: Complexity on the Brane},''
  \href{http://arxiv.org/abs/2010.16398}{{\ttfamily arXiv:2010.16398
  [hep-th]}}.

\bibitem{Matsuo:2020ypv}
Y.~Matsuo, ``{Islands and stretched horizon},''
  \href{http://arxiv.org/abs/2011.08814}{{\ttfamily arXiv:2011.08814
  [hep-th]}}.

\bibitem{Goto:2020wnk}
K.~Goto, T.~Hartman, and A.~Tajdini, ``{Replica wormholes for an evaporating 2D
  black hole},'' \href{http://arxiv.org/abs/2011.09043}{{\ttfamily
  arXiv:2011.09043 [hep-th]}}.

\bibitem{Akal:2020twv}
I.~Akal, Y.~Kusuki, N.~Shiba, T.~Takayanagi, and Z.~Wei, ``{Entanglement
  entropy in holographic moving mirror and Page curve},''
  \href{http://arxiv.org/abs/2011.12005}{{\ttfamily arXiv:2011.12005
  [hep-th]}}.

\bibitem{Basak:2020aaa}
J.~K. Basak, D.~Basu, V.~Malvimat, H.~Parihar, and G.~Sengupta, ``{Islands for
  Entanglement Negativity},'' \href{http://arxiv.org/abs/2012.03983}{{\ttfamily
  arXiv:2012.03983 [hep-th]}}.

\bibitem{Caceres:2020jcn}
E.~Caceres, A.~Kundu, A.~K. Patra, and S.~Shashi, ``{Warped Information and
  Entanglement Islands in AdS/WCFT},''
  \href{http://arxiv.org/abs/2012.05425}{{\ttfamily arXiv:2012.05425
  [hep-th]}}.

\bibitem{Raju:2020smc}
S.~Raju, ``{Lessons from the Information Paradox},''
  \href{http://arxiv.org/abs/2012.05770}{{\ttfamily arXiv:2012.05770
  [hep-th]}}.

\bibitem{Manu:2020tty}
A.~Manu, K.~Narayan, and P.~Paul, ``{Cosmological singularities, entanglement
  and quantum extremal surfaces},''
  \href{http://arxiv.org/abs/2012.07351}{{\ttfamily arXiv:2012.07351
  [hep-th]}}.

\bibitem{Deng:2020ent}
F.~Deng, J.~Chu, and Y.~Zhou, ``{Defect extremal surface as the holographic
  counterpart of Island formula},''
  \href{http://arxiv.org/abs/2012.07612}{{\ttfamily arXiv:2012.07612
  [hep-th]}}.

\bibitem{Dvali:2012en}
G.~Dvali and C.~Gomez, ``{Black Holes as Critical Point of Quantum Phase
  Transition},'' \href{http://dx.doi.org/10.1140/epjc/s10052-014-2752-3}{{\em
  Eur. Phys. J. C} {\bfseries 74} (2014) 2752},
  \href{http://arxiv.org/abs/1207.4059}{{\ttfamily arXiv:1207.4059 [hep-th]}}.

\bibitem{Dvali:2013lva}
G.~Dvali and C.~Gomez, ``{Black Hole's Information Group},''
  \href{http://arxiv.org/abs/1307.7630}{{\ttfamily arXiv:1307.7630 [hep-th]}}.

\bibitem{Dvali:2013eja}
G.~Dvali and C.~Gomez, ``{Quantum Compositeness of Gravity: Black Holes, AdS
  and Inflation},'' \href{http://dx.doi.org/10.1088/1475-7516/2014/01/023}{{\em
  JCAP} {\bfseries 01} (2014) 023},
  \href{http://arxiv.org/abs/1312.4795}{{\ttfamily arXiv:1312.4795 [hep-th]}}.

\bibitem{Ryu:2006bv}
S.~Ryu and T.~Takayanagi, ``{Holographic derivation of entanglement entropy
  from AdS/CFT},'' \href{http://dx.doi.org/10.1103/PhysRevLett.96.181602}{{\em
  Phys. Rev. Lett.} {\bfseries 96} (2006) 181602},
  \href{http://arxiv.org/abs/hep-th/0603001}{{\ttfamily arXiv:hep-th/0603001}}.

\bibitem{Hubeny:2007xt}
V.~E. Hubeny, M.~Rangamani, and T.~Takayanagi, ``{A Covariant holographic
  entanglement entropy proposal},''
  \href{http://dx.doi.org/10.1088/1126-6708/2007/07/062}{{\em JHEP} {\bfseries
  07} (2007) 062}, \href{http://arxiv.org/abs/0705.0016}{{\ttfamily
  arXiv:0705.0016 [hep-th]}}.

\bibitem{Lewkowycz:2013nqa}
A.~Lewkowycz and J.~Maldacena, ``{Generalized gravitational entropy},''
  \href{http://dx.doi.org/10.1007/JHEP08(2013)090}{{\em JHEP} {\bfseries 08}
  (2013) 090}, \href{http://arxiv.org/abs/1304.4926}{{\ttfamily arXiv:1304.4926
  [hep-th]}}.

\bibitem{Barrella:2013wja}
T.~Barrella, X.~Dong, S.~A. Hartnoll, and V.~L. Martin, ``{Holographic
  entanglement beyond classical gravity},''
  \href{http://dx.doi.org/10.1007/JHEP09(2013)109}{{\em JHEP} {\bfseries 09}
  (2013) 109}, \href{http://arxiv.org/abs/1306.4682}{{\ttfamily arXiv:1306.4682
  [hep-th]}}.

\bibitem{Faulkner:2013ana}
T.~Faulkner, A.~Lewkowycz, and J.~Maldacena, ``{Quantum corrections to
  holographic entanglement entropy},''
  \href{http://dx.doi.org/10.1007/JHEP11(2013)074}{{\em JHEP} {\bfseries 11}
  (2013) 074}, \href{http://arxiv.org/abs/1307.2892}{{\ttfamily arXiv:1307.2892
  [hep-th]}}.

\bibitem{Engelhardt:2014gca}
N.~Engelhardt and A.~C. Wall, ``{Quantum Extremal Surfaces: Holographic
  Entanglement Entropy beyond the Classical Regime},''
  \href{http://dx.doi.org/10.1007/JHEP01(2015)073}{{\em JHEP} {\bfseries 01}
  (2015) 073}, \href{http://arxiv.org/abs/1408.3203}{{\ttfamily arXiv:1408.3203
  [hep-th]}}.

\bibitem{Witten:1991yr}
E.~Witten, ``{On string theory and black holes},''
  \href{http://dx.doi.org/10.1103/PhysRevD.44.314}{{\em Phys. Rev. D}
  {\bfseries 44} (1991) 314--324}.

\bibitem{Giudice1}
G.~F. Giudice and M.~McCullough, ``{A Clockwork Theory},''
  \href{http://dx.doi.org/10.1007/JHEP02(2017)036}{{\em JHEP} {\bfseries 02}
  (2017) 036}, \href{http://arxiv.org/abs/1610.07962}{{\ttfamily
  arXiv:1610.07962 [hep-ph]}}.

\bibitem{Giudice2}
G.~F. Giudice, Y.~Kats, M.~McCullough, R.~Torre, and A.~Urbano,
  ``{Clockwork/linear dilaton: structure and phenomenology},''
  \href{http://dx.doi.org/10.1007/JHEP06(2018)009}{{\em JHEP} {\bfseries 06}
  (2018) 009}, \href{http://arxiv.org/abs/1711.08437}{{\ttfamily
  arXiv:1711.08437 [hep-ph]}}.

\bibitem{KR1}
A.~Kehagias and A.~Riotto, ``{Clockwork Inflation},''
  \href{http://dx.doi.org/10.1016/j.physletb.2017.01.042}{{\em Phys. Lett. B}
  {\bfseries 767} (2017) 73--80},
  \href{http://arxiv.org/abs/1611.03316}{{\ttfamily arXiv:1611.03316
  [hep-ph]}}.

\bibitem{KR2}
A.~Kehagias and A.~Riotto, ``{The Clockwork Supergravity},''
  \href{http://dx.doi.org/10.1007/JHEP02(2018)160}{{\em JHEP} {\bfseries 02}
  (2018) 160}, \href{http://arxiv.org/abs/1710.04175}{{\ttfamily
  arXiv:1710.04175 [hep-th]}}.

\bibitem{Laddha:2020kvp}
A.~Laddha, S.~G. Prabhu, S.~Raju, and P.~Shrivastava, ``{The Holographic Nature
  of Null Infinity},'' \href{http://arxiv.org/abs/2002.02448}{{\ttfamily
  arXiv:2002.02448 [hep-th]}}.

\bibitem{Fiola:1994ir}
T.~M. Fiola, J.~Preskill, A.~Strominger, and S.~P. Trivedi, ``{Black hole
  thermodynamics and information loss in two-dimensions},''
  \href{http://dx.doi.org/10.1103/PhysRevD.50.3987}{{\em Phys. Rev. D}
  {\bfseries 50} (1994) 3987--4014},
  \href{http://arxiv.org/abs/hep-th/9403137}{{\ttfamily arXiv:hep-th/9403137}}.

\bibitem{Huerta1}
H.~Casini and M.~Huerta, ``{A Finite entanglement entropy and the c-theorem},''
  \href{http://dx.doi.org/10.1016/j.physletb.2004.08.072}{{\em Phys. Lett. B}
  {\bfseries 600} (2004) 142--150},
  \href{http://arxiv.org/abs/hep-th/0405111}{{\ttfamily arXiv:hep-th/0405111}}.

\bibitem{Calabrese:2009ez}
P.~Calabrese, J.~Cardy, and E.~Tonni, ``{Entanglement entropy of two disjoint
  intervals in conformal field theory},''
  \href{http://dx.doi.org/10.1088/1742-5468/2009/11/P11001}{{\em J. Stat.
  Mech.} {\bfseries 0911} (2009) P11001},
  \href{http://arxiv.org/abs/0905.2069}{{\ttfamily arXiv:0905.2069 [hep-th]}}.

\bibitem{Calabrese:2010he}
P.~Calabrese, J.~Cardy, and E.~Tonni, ``{Entanglement entropy of two disjoint
  intervals in conformal field theory II},''
  \href{http://dx.doi.org/10.1088/1742-5468/2011/01/P01021}{{\em J. Stat.
  Mech.} {\bfseries 1101} (2011) P01021},
  \href{http://arxiv.org/abs/1011.5482}{{\ttfamily arXiv:1011.5482 [hep-th]}}.

\bibitem{Rabino}
S.~Elitzur, A.~Forge, and E.~Rabinovici, ``{Some global aspects of string
  compactifications},''
  \href{http://dx.doi.org/10.1016/0550-3213(91)90073-7}{{\em Nucl. Phys. B}
  {\bfseries 359} (1991) 581--610}.

\bibitem{BH2}
R.~Brustein, D.~Gorbonos, and M.~Hadad, ``{Wald's entropy is equal to a quarter
  of the horizon area in units of the effective gravitational coupling},''
  \href{http://dx.doi.org/10.1103/PhysRevD.79.044025}{{\em Phys. Rev. D}
  {\bfseries 79} (2009) 044025},
  \href{http://arxiv.org/abs/0712.3206}{{\ttfamily arXiv:0712.3206 [hep-th]}}.

\bibitem{Chow}
B.~Chow and D.~Knopf, {\em {The Ricci Flow: An Introduction (Mathematical
  Surveys and Monographs)}}.
\newblock American Mathematical Society; UK ed., 2004.

\bibitem{Bakas}
I.~Bakas, ``{Renormalization group equations and geometric flows},'' {\em Ann.
  U. Craiova Phys.} {\bfseries 16} no.~II, (2006) 20--29,
  \href{http://arxiv.org/abs/hep-th/0702034}{{\ttfamily arXiv:hep-th/0702034}}.

\bibitem{KLL1}
A.~Kehagias, D.~L\"ust, and S.~L\"ust, ``{Swampland, Gradient Flow and Infinite
  Distance},'' \href{http://dx.doi.org/10.1007/JHEP04(2020)170}{{\em JHEP}
  {\bfseries 04} (2020) 170}, \href{http://arxiv.org/abs/1910.00453}{{\ttfamily
  arXiv:1910.00453 [hep-th]}}.

\bibitem{Huerta2}
H.~Casini and M.~Huerta, ``{A c-theorem for the entanglement entropy},''
  \href{http://dx.doi.org/10.1088/1751-8113/40/25/S57}{{\em J. Phys. A}
  {\bfseries 40} (2007) 7031--7036},
  \href{http://arxiv.org/abs/cond-mat/0610375}{{\ttfamily
  arXiv:cond-mat/0610375}}.

\bibitem{Dimitris}
C.-S. Chu and D.~Giataganas, ``{$c$-Theorem for Anisotropic RG Flows from
  Holographic Entanglement Entropy},''
  \href{http://dx.doi.org/10.1103/PhysRevD.101.046007}{{\em Phys. Rev. D}
  {\bfseries 101} no.~4, (2020) 046007},
  \href{http://arxiv.org/abs/1906.09620}{{\ttfamily arXiv:1906.09620
  [hep-th]}}.

\end{thebibliography}\endgroup

\end{document}